\documentclass[9pt,twocolumn,twoside,lineno]{pbu-new}
\pdfoutput=1
\templatetype{pnasresearcharticle} 

\title{Persistent structures in a 3D dynamical system with solid and fluid regions}

\author[a]{Zafir Zaman}
\author[a]{Mengqi Yu}
\author[b]{Paul P. Park}
\author[a,c,d]{Julio M. Ottino}
\author[c,a,d]{Richard M. Lueptow}
\author[c]{Paul B. Umbanhowar}

\affil[a]{Department of Chemical and Biological Engineering, Northwestern University, Evanston, IL 60208 USA}
\affil[b]{Department of Engineering Sciences and Applied Mathematics, Northwestern University, Evanston, IL 60208 USA}
\affil[c]{Department of Mechanical Engineering, Northwestern University, Evanston, IL 60208 USA}
\affil[d]{The Northwestern Institute on Complex Systems (NICO), Northwestern University, Evanston, IL 60208 USA}

\leadauthor{Zaman}

\correspondingauthor{\textsuperscript{1}To whom correspondence should be addressed. E-mail: umbanhowar@northwestern.edu}


\begin{abstract}

Remarkably persistent mixing and non-mixing regions (islands) are observed to coexist in a three-dimensional dynamical system where randomness is expected. The track of an x-ray opaque particle in a spherical shell half-filled with dry non-cohesive particles and periodically rotated about two axes reveals interspersed structures that are spatially complex and vary non-trivially with the rotation angles. The geometric skeleton of the structures forms from the subtle interplay between fluid-like mixing by stretching-and-folding, and solids mixing by cutting-and-shuffling, which is described by the mathematics of piecewise isometries. In the physical system, larger islands predicted by the cutting-and-shuffling model alone can persist despite the presence of stretching-and-folding flows and particle-collision-driven diffusion, while predicted smaller islands are not observed. By uncovering the synergy of simultaneous fluid and solid mixing, we point the way to a more fundamental understanding of advection driven mixing in materials with both solid and flowing regions.
\end{abstract}

\begin{document}

\verticaladjustment{-2pt}

\maketitle
\thispagestyle{firststyle}
\ifthenelse{\boolean{shortarticle}}{\ifthenelse{\boolean{singlecolumn}}{\abscontentformatted}{\abscontent}}{}

\dropcap{T}he goal of mixing is to rearrange initially segregated matter into states where the constituent elements are homogeneously distributed.  In fluids, where the elements are atoms or molecules, mixing at low Reynolds numbers can be achieved by the stretching-and-folding of chaotic flows combined with thermal diffusion which drives mixing at the smallest length scales. In bulk solids composed of macroscopic (athermal) particles, the particles can be deliberately rearranged, as in the cutting-and-shuffling of a deck of cards. Other rheological materials--Bingham fluids for example--fall between these two extremes. While mixing of fluids and mixing of solids have long histories and are relatively mature fields, little is understood about mixing in three-dimensions (3D) when solid and flowing regions coexist. For example, in yield stress materials constituent elements move together as a solid where local stresses are low, but flow in relative motion where the yield stress is exceeded.  Common examples of yield stress materials include paint, wet concrete, polymer mixtures,\cite{ZhouMacoskoJRheol2002} and granular materials, e.g., sand.  Understanding 3D mixing in such materials is critical in many domains where variations in local concentration can be disastrous, including the pharmaceutical industry,\cite{MuzzioPowderTech2002} composite materials,\cite{UpadhyayaPMTech1997} and concrete manufacturing. Also noteworthy is the lottery with the largest payout in the world, the Spanish Christmas Lottery, where 100,000 wooden balls are mixed by rotating a 2\,m diameter sphere about a horizontal axis with the expectation of randomness. Our results show that this expectation cannot be taken for granted.

\begin{figure}[t!]
	\includegraphics{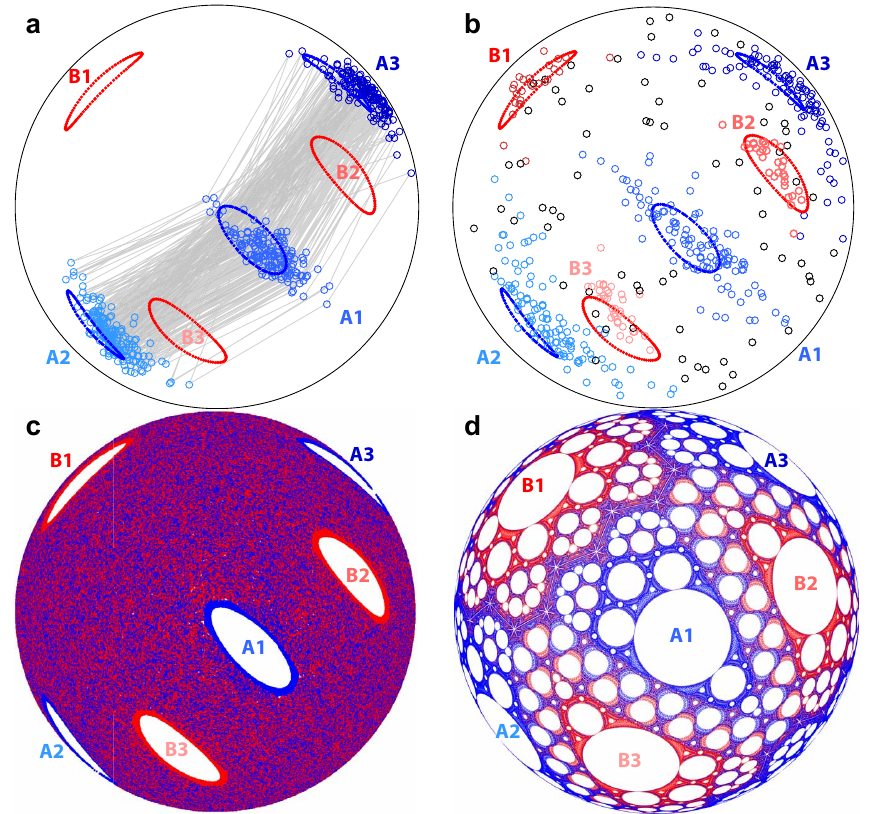}
	\caption[Period-3 non-mixing regions under the $(57^\mathrm{o}, 57^\mathrm{o})$ protocol]{\label{fig:protocol5757} \textbf{Period-3 non-mixing regions under the $(57^\mathrm{o}, 57^\mathrm{o})$ protocol (bottom views).} \textbf{a}, Tracer particle cycles between period-3 regions A1-A3 (blue circles) with stroboscopic paths (gray lines) in a 500-iteration experiment. \textbf{b}, In a second experiment, tracer particle cycles between period-3 regions A1-A3 (blue circles) for 326 iterations (1-90, 96-331), period-3 region B1-B3 (red circles) for 101 iterations (386-486), and outside of period-3 regions (black circles) for 73 iterations (91-95, 332-385, and 487-500). \textbf{c}, Passive tracers in finite flowing layer (FL) model mix everywhere except in two sets of period-3 regions (A1-A3 and B1-B3). \textbf{d}, Piecewise isometry model--which describes solids mixing by cutting-and-shuffling--predicts unmixed cells (white regions) of which only the largest cells, (A1-A3, B1-B3) persist in the flowing layer model (\textbf{c}).  Period-3 trapping regions from experiments correspond with period-3 cells from the FL and PWI models [blue (red) curves for regions A1-A3 (B1-B3)].}
\end{figure}

To study the interaction of mixing by stretching-and-folding with mixing by cutting-and-shuffling, we consider a geometrically simple 3D system with localized flow --- a spherical tumbler half-filled with a dry granular material and rotated alternately about orthogonal horizontal axes by angles $(\theta_z,\theta_x)$ beyond the repose angle of the granular material, $\beta$. In tumblers, particles flow in a relatively thin layer at the free surface, while below the surface, particles move together in solid body rotation about the rotation axis.  Using X-ray imaging, we mapped the location of a 4\,mm diameter spherical tracer particle in a $D(=2R)=14$\,cm diameter spherical tumbler half filled with 2\,mm diameter glass spheres after each iteration for numerous protocols [i.e., $(\theta_z,\theta_x)$ pairs]. The larger diameter tracer particle flows on the free surface and is subsequently deposited in the bed near the tumbler wall.

\subsection*{Results and Discussion}

Consider first the itinerary of a tracer particle during 500 iterations of the ($57^\mathrm{o}, 57^\mathrm{o}$) protocol, see Fig.~\ref{fig:protocol5757}(a). After each iteration the tracer particle was alternately displaced between three distinct regions as indicated by the gray lines. Figure~\ref{fig:protocol5757}(b) shows a second 500 iteration experiment in which the tracer particle moved periodically between the same three regions (blue points) and between a second set of period-3 regions (red points) as well as aperiodically (black points). Similar period-3 non-mixing regions are observed in the ``nearby'' protocols ($54^\mathrm{o}, 54^\mathrm{o}$) and ($60^\mathrm{o},60^\mathrm{o}$).

To better understand the existence of these non-mixing and mixing regions, we consider the predictions of a standard advection based continuum model~\cite{MeierAdvPhys2007} for tumbler flow, in which, for rotation about any axis (the $z$-axis here with $x$ in the streamwise direction and $y$ normal to the free surface), the non-dimensionalized velocity field $\mathbf{u} = (u,v,w)$ is piecewise defined such that the flowing layer $(0\geq y\geq-\delta)$ velocity is $\mathbf{u}_{\mathrm{fl}} = ((\delta+y)/\epsilon^2,xy/\delta,0)$ and the bulk $(y<-\delta)$ solid body rotation velocity is $\mathbf{u}_{\mathrm{b}} = (y,-x,0).$ The lenticular flowing layer interface with the bulk is located at $\delta (x,z) = \epsilon \sqrt{1-x^2-z^2}$, where $\epsilon = \delta(0,0) = \sqrt{\omega/\dot{\gamma}}$ is the maximal dimensionless flowing layer thickness at the center of the sphere ($x=z=0$) for shear rate $\dot{\gamma}$ and angular rotation velocity $\omega$.  All variables are dimensionless---lengths ($x$, $y$, $z$, $r$, $\delta$) are normalized by the tumbler radius $R$ and rotation period $T$ is normalized by $1/\omega$. This flowing layer (FL) model, which includes stretching characteristic of chaotic flows\cite{Ottino1989Book}, is parametrized by $\epsilon$, which is set to 0.15 based on the particles used in the experiments (see Methods section). To characterize mixing in the model, blue passive tracer points are seeded at the intersection of the flowing layer boundary $\delta$ with a hemispherical subshell having normalized radius $r=0.9$\footnote{\label{fnote1}Radius $r = 0.9$ was used to mimic tracer particle position in experiments instead of a larger radius ($0.95\leq r\leq0.97$) as elliptic regions at larger $r>0.9$ reveal more intricate structures that are not apparent in experiments likely due to particle size effects and collisional diffusion. As the intricate elliptic regions at $r>0.9$ occur at the same location on the hemispherical shell as elliptic regions at $r=0.9$, we use the elliptic orbits at $r=0.9$ for comparison to experiment for visual clarity.} before the first rotation, while red points are seeded in the same location after a half iteration (i.e., rotation about the $z$-axis only). In Fig.~\ref{fig:protocol5757}(c), the Poincar\'e (stroboscopic) map of tracer points advected by the model displays uniform mixing throughout the domain except for six empty regions (A1-A3 and B1-B3). These non-mixing regions correspond to the regions in experiments where the tracer particle lingers, as indicated by their mapping onto Figs.~\ref{fig:protocol5757}(a,b).

Although the FL model captures the main features of the experiments shown in Figs.~\ref{fig:protocol5757}(a,b), deeper insight into why persistent structures form is gained by focusing exclusively on solids mixing. This is accomplished by taking the $\delta\to0$ limit (an infinitely thin flowing layer) in which tracers instantaneously jump from the point they reach the free surface to a downstream point symmetric about the midpoint of the free surface and, consequently, undergo only solid body rotation. Since there is no flowing layer and, consequently, no shear, the biaxial mixing protocol in the $\delta\to0$ limit corresponds to a radially invariant hemispherical domain with mixing dynamics that are equivalent to slicing the hemisphere into pieces that are rearranged and then reassembled into a hemisphere again. This type of transformation is called a piecewise isometry (PWI).\cite{GoetzPWI2003} Similar to the FL model, the boundaries formed by the slicing after a rotation about a single axis are used as initial conditions for the tracer points whose trajectories form a subset of the exceptional set,\cite{FuDuanPhysicaD2008} which is the skeleton for the transport dynamics in PWI systems. Here, blue tracers are seeded where the domain is cut by the action of the $\theta_z$ rotation, while red tracers are seeded where the domain is cut by the $\theta_x$ rotation, similar to the initial conditions used in the FL model. The mixing mechanism for a piecewise isometry is simply cutting-and-shuffling in analogy with mixing a deck of cards.\cite{TrefethenPRSA2000}

Similar to the FL model, iteration of the PWI model for the ($57^\mathrm{o}, 57^\mathrm{o}$) protocol (Fig.~\ref{fig:protocol5757}(d)) generates open regions avoided by the tracers, known as cells. These cells vary in size and periodicity across the domain, with certain areas dominated by a particular color of tracer particle. The largest circular cells have the lowest periodicity.  The similarity in size and location between the largest cells in the PWI model, the elliptical domains in the FL model (Fig.~\ref{fig:protocol5757}(c)), and the non-mixing regions in the experiments (Figs.~\ref{fig:protocol5757}(a,b)) is clear. This agreement suggests that the periodic regions observed in experiment for the ($57^\mathrm{o}, 57^\mathrm{o}$) protocol result from the solids mixing structure generated by cutting-and-shuffling.

\begin{figure}[t!]
	\includegraphics{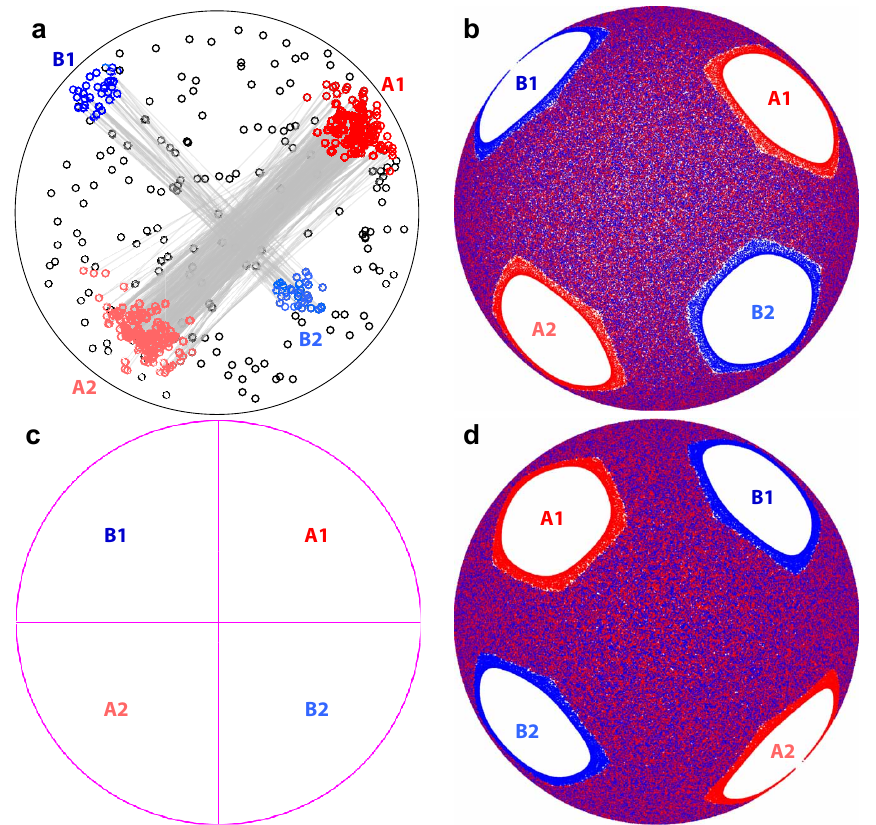}
	\caption[Period-2 non-mixing regions in the $(90^\mathrm{o}, 90^\mathrm{o})$ protocol]{\label{fig:protocol9090} \textbf{Period-2 non-mixing regions under the $(90^\mathrm{o}, 90^\mathrm{o})$ protocol.} \textbf{a}, The tracer particle in a 500-iteration experiment is confined to period-2 regions A1-A2 (red circles) for 263 iterations (mean 37.6 iterations per instance) and B1-B2 (blue circles) for 46 iterations (mean 11.5 iterations per instance), stroboscopic paths indicated by gray lines. \textbf{b}, In the FL model, two period-2 (A1-A2 and B1-B2) regions exist. \textbf{c}, The PWI model undergoes trivial period-2 rearrangement of the entire domain. \textbf{d}, Interchange of non-mixing A and B regions in the FL model at 500 plus one-half iteration illustrates the generic half-period offset relationship between A and B non-mixing regions for orthogonal rotation axes.}
\end{figure}

The periodicity of non-mixing structures depends on the protocol, for example consider the results for the ($90^\mathrm{o}, 90^\mathrm{o}$) protocol shown in Fig.~\ref{fig:protocol9090}.  In experiment (Fig.~\ref{fig:protocol9090}(a)), a tracer particle initially seeded in the A1 region, cycles between the A1 and A2 regions and between the B1 and B2 regions for a total of 263 and 46 iterations, respectively. Compared to the period-3 non-mixing regions under the ($57^\mathrm{o}, 57^\mathrm{o}$) protocol, the tracer escapes more frequently from the period-2 non-mixing regions. The non-mixing regions A1-A2 and B1-B2 correspond to the white regions avoided by the tracer points in the FL model (Fig.~\ref{fig:protocol9090}(b)), while in the PWI model (Fig.~\ref{fig:protocol9090}(c)) non-mixing regions occupy the entire domain since the entire hemisphere returns to its initial condition every two iterations. As for the ($57^\mathrm{o}, 57^\mathrm{o}$) protocol, the finite-depth flowing layer converts a portion of the PWI model's non-mixing regions into mixing regions.

For the half-full spherical tumbler mixed by alternating rotations about orthogonal axis used here, persistent non-mixing period-n regions always appear in pairs, a manifestation of the relationship between the full and half iteration structures. To demonstrate, the sets of non-mixing regions shown in Fig.~\ref{fig:protocol9090}(b) swap positions upon an additional half-cycle of the mixing protocol (Fig.~\ref{fig:protocol9090}(d)). Thus, the B-set of non-mixing regions is the half-period offset of the A-set of non-mixing regions.

Both PWI and FL models capture the large scale persistent mixing and non-mixing structures observed in experiments for the protocols we have examined so far, but they do not describe the transitions of the tracer particle between islands seen in experiments, e.g., Fig.~\ref{fig:protocol5757}(b) and Fig.~\ref{fig:protocol9090}(a).  This is because neither model includes diffusion, which in granular flows is driven by particle-particle and particle-wall collisions. The collisional diffusion coefficient in flowing macroscopic particles scales as $d^2\dot{\gamma}$ where $d$ is the diameter of the particles in the tumbler and $\dot{\gamma}$ is the shear rate\cite{UtterPhysRevE2004}; in our tumbler, diffusion displaces a particle on average by about its diameter per flowing layer pass.\cite{ZamanPRE2013} For the ($57^\mathrm{o}, 57^\mathrm{o}$) protocol which produces $\sim1$ flowing layer pass per iteration, the root-mean-square displacement after 500-iterations ($d\sqrt{500}$) is 4.5\,cm (approximately the tumbler radius). Based on the size of the non-mixing regions under the ($57^\mathrm{o}, 57^\mathrm{o}$) protocol, we expect a mean residency time of about 350 iterations in the regions, which is on the order of the observed residence times of at least 500 iterations in Fig.~\ref{fig:protocol5757}(a), 326 iterations for the period-3 A regions in Fig.~\ref{fig:protocol5757}(d), and 101 iterations for the period-3 B regions in in Fig.~\ref{fig:protocol5757}(d). Consequently, particles in experiments ``leak'' out of non-mixing regions with a residence time that decreases as the square root of the size of the non-mixing regions.

\begin{figure}[b!]
	\includegraphics{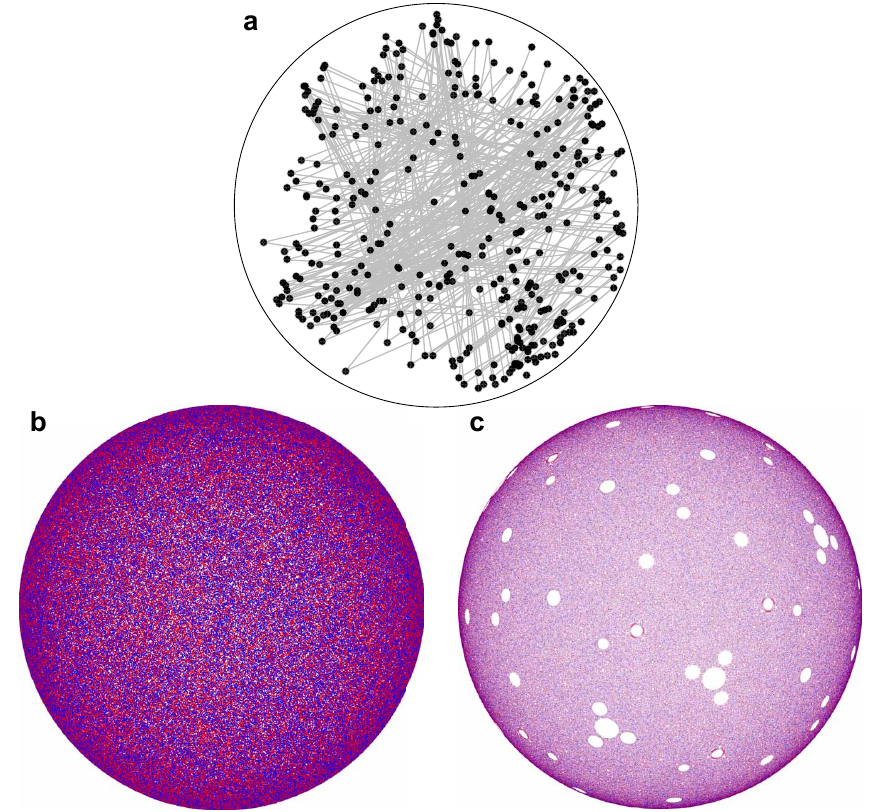}
	\caption[Mixing for the $(75^\mathrm{o}, 60^\mathrm{o})$ protocol]{\label{fig:protocol7560} \textbf{Mixing under the $(75^\mathrm{o}, 60^\mathrm{o})$ protocol.} \textbf{a}, Tracer particle positions (black points) and stroboscopic paths (gray lines) in a 500-iteration experiment continuously explore nearly the entire domain with no obvious non-mixing regions. \textbf{b}, In the corresponding FL model, tracer points are mixed throughout the domain, consistent with the experiment. \textbf{c}, The PWI model predicts a mostly mixed domain with a few small cells that are unobserved in the experiment or FL model.}
\end{figure}

Other protocols can produce completely mixed domains. For example under the rotationally asymmetric ($75^\mathrm{o}, 60^\mathrm{o}$) protocol (Fig.~\ref{fig:protocol7560}), the tracer particle explores most of the domain in experiments (Fig.~\ref{fig:protocol7560}(a)), while the FL (Fig.~\ref{fig:protocol7560}(b)) and PWI (Fig.~\ref{fig:protocol7560}(c)) models generate mixing regions that completely and nearly completely, respectively, fill the domain. Mixing is not guaranteed by an asymmetric protocol, i.e., $\theta_z\neq\theta_x$ (see Extended Data Fig.~1(a)), and, conversely, a symmetric protocol, i.e., $\theta_z=\theta_x$, does not guarantee the existence of non-mixing regions.

\begin{figure}[b!]
	\includegraphics{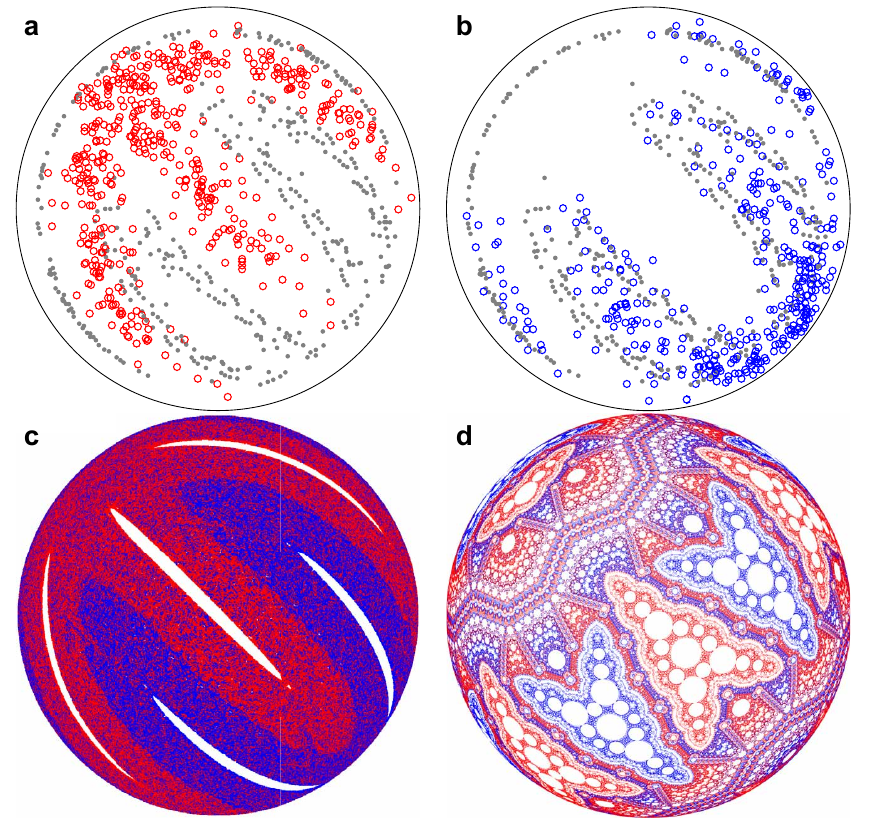}
	\caption[Period-3 frustrated mixing between two halves of the domain for the $(45^\mathrm{o}, 45^\mathrm{o})$ protocol]{\label{fig:protocol4545} \textbf{Period-3 frustrated mixing between two halves of the domain in the $(45^\mathrm{o}, 45^\mathrm{o})$ protocol.} \textbf{a-b}, In 500-iteration experiments, the tracer particle (red and blue circles) cycles through two distinct parts of the domain every three periods depending upon the initial condition. \textbf{c}, The corresponding FL model depicts two thin sets of period-3 islands corresponding to the structure in experiment \textbf{a-b} (red islands in \textbf{c} correspond to red ``fingers'' in \textbf{a} and blue islands in \textbf{c} correspond to blue ``fingers'' in \textbf{b}). The two sets of islands are separated by a mixing barrier (grey points in \textbf{a-b}) extracted from the FL model (\textbf{c}). \textbf{d}, The PWI model predicts an arrowhead pattern of cells.}
\end{figure}

Under the three protocols shown in Figs.~\ref{fig:protocol5757}-\ref{fig:protocol7560}, when non-mixing regions appear in experiments, they are present in the FL model and correspond with the largest cells in the PWI model. In contrast, under the ($45^\mathrm{o}, 45^\mathrm{o}$) protocol (Fig.~\ref{fig:protocol4545}), a mixing barrier emerges that is not captured by the PWI model. A tracer particle in two experiments (Fig.~\ref{fig:protocol4545}(a-b)) follows two different extended finger-like period-3 structures that are separated by a leaky barrier to particle transport and together cover the entire domain. These interdigitated period-3 structures are readily apparent in the FL model (Fig.~\ref{fig:protocol4545}(c)). Red and blue tracers each dominate half of the domain with an elongated non-mixing region in each ``finger,'' which roughly corresponds to the tracer particle positions from experiment. The region dominated by the blue tracers appears to have only two ``fingers'', because the third ``finger'' is mostly contained in the flowing layer, which is not visible in the view shown in Fig.~\ref{fig:protocol4545}(c). The blue edges of this ``finger'' are evident at the periphery of the domain as well as the periphery in the experiments (Fig.~\ref{fig:protocol4545}(b)). For the equivalent half iteration structures in the FL model, the red tracer dominated regions alternate with the regions dominated by blue tracers with one of the red ``fingers'' mapped to the flowing layer. To delineate the two regions, we followed a tracer point in the FL model located between the red and blue tracer dominated regions to produce the grey points in Fig.~\ref{fig:protocol4545}(a-b). The path of this tracer suggests a mixing barrier that wraps around the entire domain.

Unlike the PWI model predictions for protocols shown in Figs.~\ref{fig:protocol5757}-\ref{fig:protocol7560}, the predicted structure under the $(45^\mathrm{o}, 45^\mathrm{o})$ protocol (Fig.~\ref{fig:protocol4545}(d)) is less clearly related to the experiment and the FL model as it lacks the typical large cells that manifest as non-mixing regions in the experiment. Instead the PWI model predicts large arrowhead-like features consisting of multiple small cells with transport barriers between red and blue arrowheads. The colors and positions of the arrowhead features correspond to the ``finger'' features in the experiment and FL model.

\begin{figure}[t!]
	\includegraphics{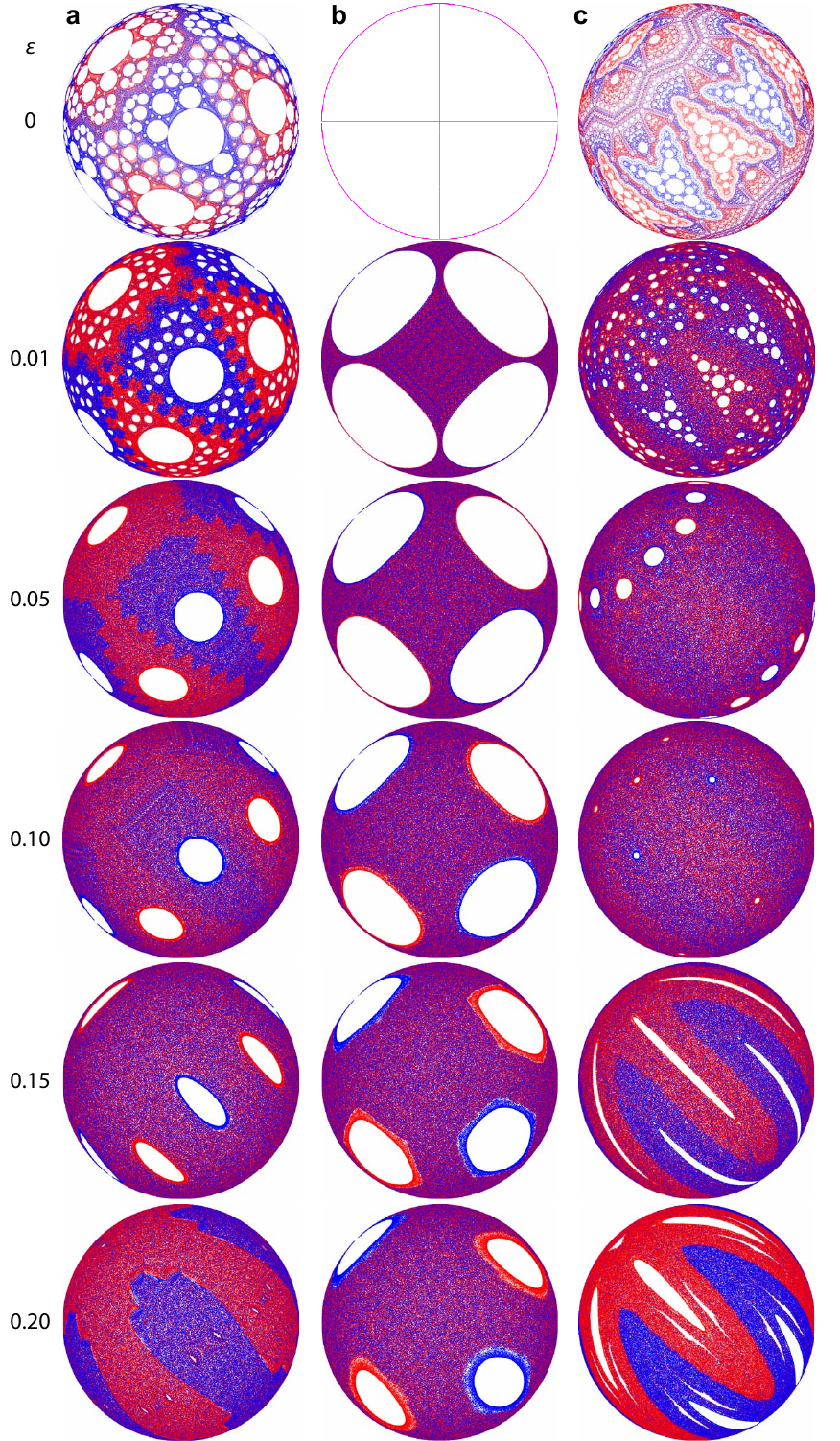}
	\caption[Effect of flowing layer thickness on non-mixing regions]{\label{fig:epsilon_var} \textbf{Effect of flowing layer thickness on non-mixing regions.} Under \textbf{a} $(57^\mathrm{o}, 57^\mathrm{o})$,  \textbf{b} $(90^\mathrm{o}, 90^\mathrm{o})$, and \textbf{c} $(45^\mathrm{o}, 45^\mathrm{o})$ protocols and with increasing flowing layer thickness, non-mixing regions formed by cutting-and-shuffling ($\epsilon=0$) shrink (\textbf{a}, \textbf{b}, \textbf{c}) and disappear (\textbf{a}, \textbf{c}), while new non-mixing regions originating from stretching-and-folding in the finite thickness flowing layer appear and grow (\textbf{c}, $\epsilon>=0.15$).}
\end{figure}

The structures predicted by the FL model are in close agreement with experimental observations regardless of the protocol, while the structures predicted by the PWI model are not always observed.  For example, the smaller non-mixing cells in the PWI model for the four protocols we examine above are not seen in experiments.  To better understand the relationship between the non-mixing structures predicted by the PWI and FL models, particularly for the ($45^\mathrm{o}, 45^\mathrm{o}$) protocol, we consider the influence of flowing layer thickness. Physically, flowing layer thickness increases with rotation rate $\omega$ and decreases with shear rate $\dot{\gamma}$, since $\epsilon = \sqrt{\omega/\dot{\gamma}} = \delta(0,0)/R$.\cite{FelixEPJE2007} Additionally, the flowing layer is typically thicker than $\sim10$ particle diameters in experiments. As Fig.~\ref{fig:epsilon_var} shows, non-mixing regions shrink and, in most cases, vanish with increasing flowing layer thickness. All cells in the PWI model (i.e., $\epsilon = 0$) disappear under the $(57^\mathrm{o}, 57^\mathrm{o})$ and $(45^\mathrm{o}, 45^\mathrm{o})$ protocols at $\epsilon = 0.20$ and $\epsilon = 0.05$, respectively, while under the ($90^\mathrm{o}, 90^\mathrm{o}$), cells shrink but persist up to the largest $\epsilon$ examined (0.20).

Focussing on the ($45^\mathrm{o}, 45^\mathrm{o}$) protocol (Fig.~\ref{fig:epsilon_var}(c)), the cells present at $\epsilon=0$ shrink and eventually annihilate with increasing $\epsilon$. However, new structures emerge for larger $\epsilon,$ characterized by both non-mixing regions and the mixing barrier between red and blue tracer particles (similar behavior is seen for the ($45^\mathrm{o}, 15^\mathrm{o}$) protocol shown in Extended Data Fig.~1). In particular, at $\epsilon = 0.15$, the leftmost red ``finger'' and its central non-mixing region land in the flowing layer after a half iteration, while each of the subsequent red ``fingers'' map to the flowing layer one iteration apart. The same phenomenon occurs for the blue ``fingers'' except on full iterations. The flowing layer acts as a mixing barrier in the ($45^\mathrm{o}, 45^\mathrm{o}$) protocol as each part of the fixed bed lands in the flowing layer during a change in the rotation axis.

\begin{figure}[h!]
	\includegraphics{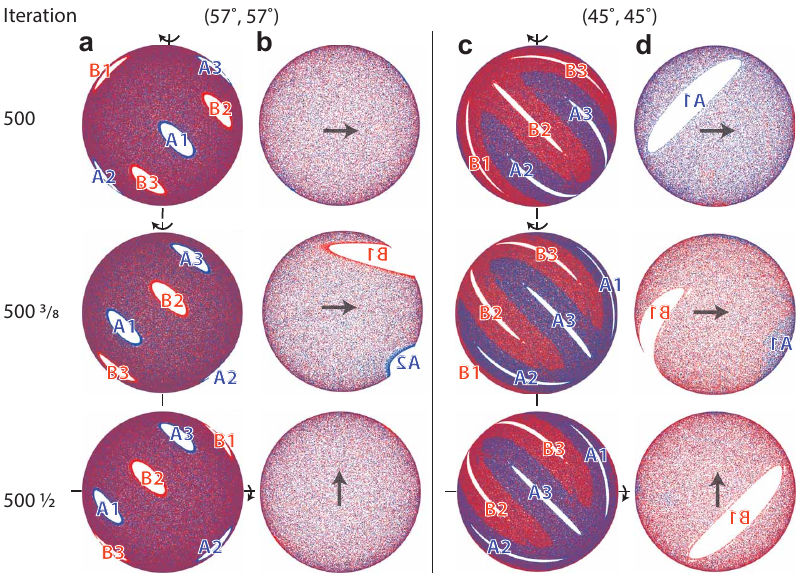}
	\caption[Motion of periodic non-mixing regions.]{\label{fig:vidframes} \textbf{Motion of periodic non-mixing regions.} At finite flowing layer thickness ($\epsilon=0.15$), \textbf{a-b}, persistent non-mixing regions (A1-A3, B1-B3) present at $\epsilon=0$ pass fully through the flowing layer after each half-iteration (($57^\mathrm{o}, 57^\mathrm{o}$) protocol), while, \textbf{c-d}, emergent non-mixing regions (A1-A3, B1-B3) that appear only for $\epsilon>0$ and their accompanying ``fingers'' alternate between being fully contained in the flowing layer or in the bulk after each half-iteration (($45^\mathrm{o}, 45^\mathrm{o}$) protocol).  Views from below of the hemispherical shell (\textbf{a} and \textbf{c}) and the planar flowing layer (\textbf{b} and \textbf{d}). Labels of periodic regions illustrate changes in orientation. Animations are provided in supplementary videos.}
\end{figure}

Differences between non-mixing regions present at $\epsilon=0$ and the non-mixing regions that can emerge for $\epsilon>0$ are illustrated in Fig.~\ref{fig:vidframes} (and in an accompanying supplemental video). Persistent non-mixing regions present at $\epsilon=0$ pass entirely through the flowing layer during each rotation, and their boundaries are set by the boundary of the flowing layer during the interchange of rotation axes, see Fig.~\ref{fig:vidframes}(a-b). As a result, the periodicity of the non-mixing region is determined by the rotation angle, and periodicity increases as rotation angle decreases because the number of flowing layer passes for a non-mixing region between actions decreases. In contrast, the non-mixing regions that appear only for $\epsilon>0$ periodically land entirely within the flowing layer at the end of a rotation. Their boundaries, unlike persistent non-mixing regions, are not necessarily coincident with the flowing layer boundary at the end of a rotation. In the example shown in Fig.~\ref{fig:vidframes}(c-d), the non-mixing regions in the bulk are long and thin and their boundaries contact the flowing layer boundary when two of the trio are in the bulk (A2 and A3 at 500 iterations, and B2 and B3 at 500-1/2 iterations) while the other region lands in the flowing layer. Additionally, the boundary of the mixing barrier between red and blue ``fingers'' maps approximately to the boundary of the flowing layer, but is less clearly defined than the non-mixing region boundaries.

Based on this study of granular material in a spherical tumbler under a bi-axial mixing protocol, we expect other 3D dynamical systems with coexisting solid and fluid regions to exhibit similar complex mixing structures that result from the distinct and competing mixing processes associated with fluid stretching-and-folding and solid cutting-and-shuffling. A deeper understanding of these \emph{hybrid-mixing} systems based on the structures determined by piecewise isometry theory is expected when the spatial extent of flowing regions is limited or when the flows are weakly shearing.

\matmethods{
\subsection*{Experiments}
Experiments were conducted using $d = 1.89 \pm 0.09$-mm diameter soda-lime glass beads (SiLiglit Deco Beads, Sigmund Lindner GmbH, Germany) in a half-filled $D=14$-cm diameter acrylic spherical tumbler. A density matched ($\rho = 2.5$ g cm$^{-1}$) $d_{\mathrm{tracer}}=4$-mm diameter X-ray opaque tracer particle constructed from two 3D-printed plastic hemispherical shells with a Pb-Sn solder sphere in the center was used to visualize the flow under X-ray imaging. The larger diameter of the tracer particle caused the particle to flow at the surface of the flowing layer and near the tumbler wall in the solid bed (bold path in Extended Data Fig.\,2). The tumbler was mounted on an apparatus that performs the biaxial protocol while being X-ray imaged. For each action ($\mathbf{U}$ or $\mathbf{W}$, in Extended Data Fig.\,2), the sphere was rotated about a single axis for protocol angle ($\theta_z$ or $\theta_x$) by three wheels driven by a motor mounted on the turntable at rotation speed $\omega = 2.6$ rpm. Since particles only flow continuously once the free surface reaches the dynamic angle of repose $\beta$ with respect to the horizontal,\footnote{Since the particle bed was stationary at the onset of flow, the free surface was at the static angle of repose $\beta_s$. There was a small avalanche at the start of flow where the free surface first relaxed to the dynamic angle of repose $\beta < \beta_s$} the tumbler was slowly rotated to the angle of repose prior to each action.  After each action, the tumbler was rotated in the reverse direction to ensure the free surface was horizontal before switching the rotation axis. The apparatus reoriented the drive wheels for the next rotation by raising the tumbler off the wheels, rotating the turntable to the next tumbling axis, and then lowering the tumbler back down onto the wheels.

\begin{figure*}[h!]
	\includegraphics[width=1\linewidth]{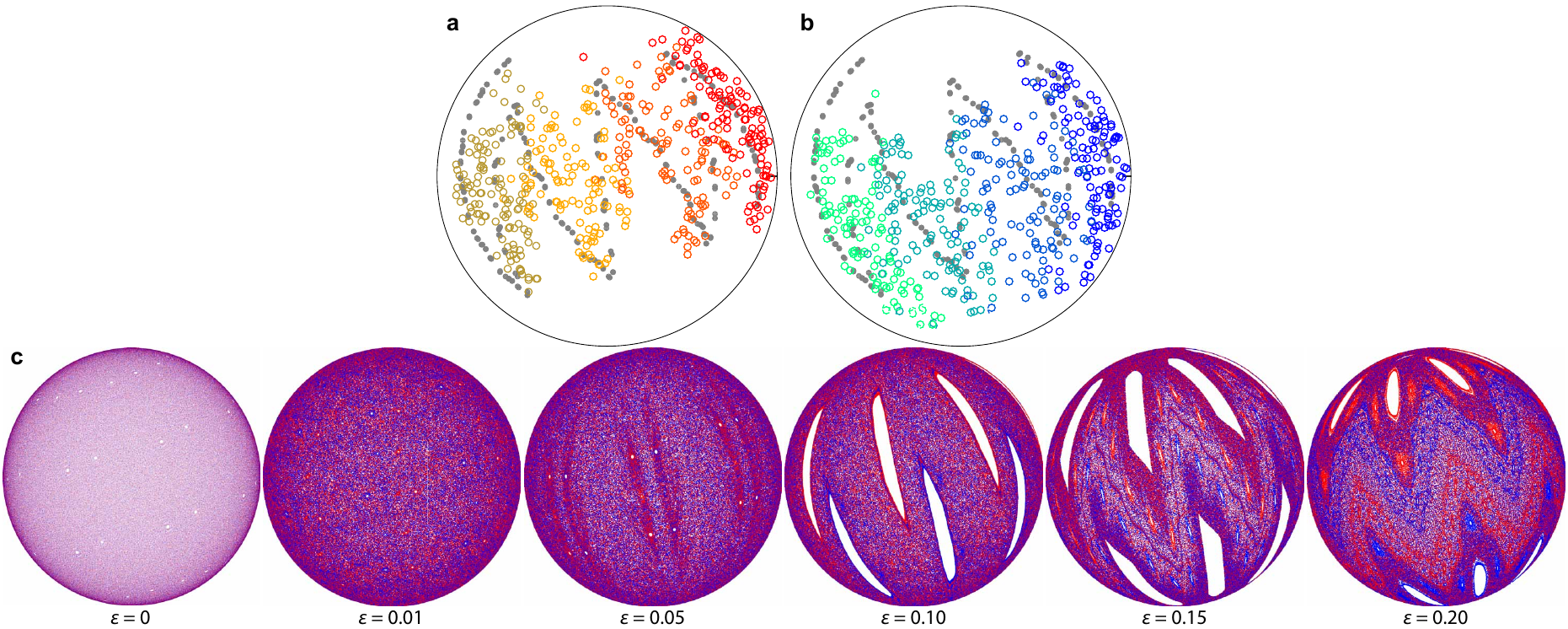}
	\caption[Weak elliptic non-mixing barrier around period-4 regions under the $(45^\mathrm{o}, 15^\mathrm{o})$ protocol]{\label{fig:protocol4515} \textbf{Extended Data Figure 1 $\vert$ Weak elliptic non-mixing barrier around period-4 regions in $(45^\mathrm{o}, 15^\mathrm{o})$ protocol.}  \textbf{a}-\textbf{b}, The tracer particle in 500-iteration experiments sometimes crosses the mixing barrier (grey points). Color map illustrates period-4 cycle in the bulk. \textbf{c}, Non-mixing regions become prominent with increasing flowing layer thickness in the FL model.}
\end{figure*}

\begin{figure*}[h!]
	\includegraphics[width=1\linewidth]{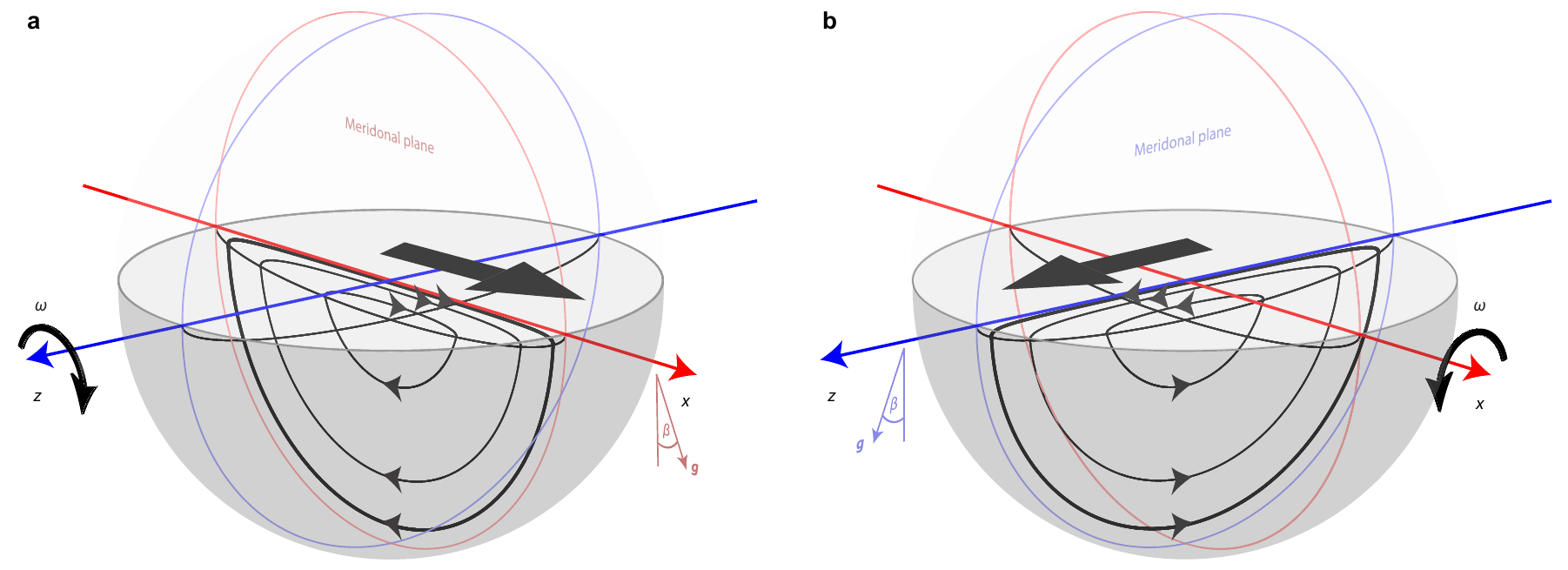}
	\caption{\label{fig:BST_schematic} \textbf{Extended Data Figure 2 $\vert$ Bi-axial spherical tumbler flow.} The flow consists of an alternating sequence of two single axis actions $\mathbf{U}$ (\textbf{a}) and $\mathbf{W}$ (\textbf{b}) about the $z$- and $x$-axis, respectively, at rotation speed $\omega$. During each action, particles form a thin lenticular flowing layer that flows in the direction normal to the rotation axis (large flat black arrow near the center of the tumbler). Particles from the downstream half of the flowing layer are deposited onto the solid bed of particles that rotates with the tumbler wall, and they remain there until they reach the flowing layer again. The angle of the flowing free surface is $\beta$ with respect to gravity, $\mathbf{g}$. Three mean particle paths are shown for a two-dimensional slice for each action, with the bold path (near the free surface and closest to tumbler wall) being a representative path for the large X-ray opaque tracer particle.}	
\end{figure*}

\begin{figure*}
	\includegraphics{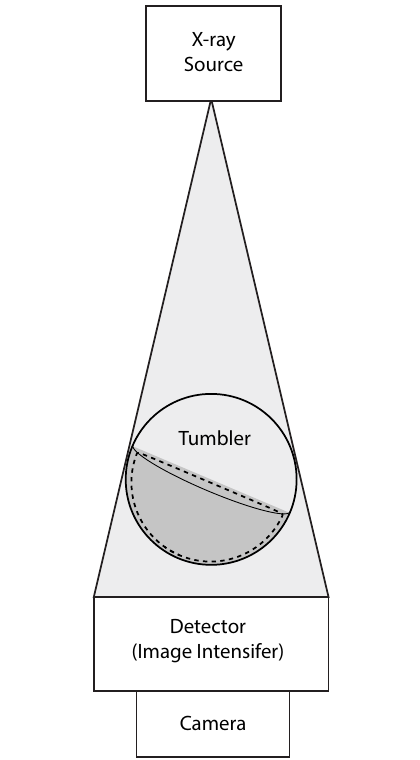}
	\caption{\label{fig:tumblersys_schematic} \textbf{Extended Data Figure 3 $\vert$ Apparatus.} Front view cross-section of the tumbler system. X-ray source projects a vertical cone beam over the spherical tumbler and X-ray image intensifier. A camera captures the resulting image. Vertical cross section of the half full tumbler shows the bed (dark gray region) with a thin lenticular flowing layer. The larger density matched X-ray opaque tracer particle path (dashed curve) is in solid body rotation close to the tumbler wall in the bed and flows along the free surface of the flowing layer.}	
\end{figure*}

To minimize accumulated error from executing the protocol over large numbers of iterations, a position sensor was used to ensure that the turntable returned to the original axis after each iteration. Additionally, a thin X-ray opaque fiducial marker (lead tape) was mounted on the interior wall of the spherical tumbler to track any tumbling deviations. With these measures in place, the system has an angular displacement error of less than $1^\mathrm{o}$ per protocol iteration. The errors are not systematic and, hence, average to zero.

Components of the spherical tumbler apparatus within the X-ray beam path consist of X-ray transparent materials (aluminum and plastic) to ensure a clear image suitable for particle tracking. Lead tape was also affixed to the turntable to reduce variations in X-ray beam attenuation due to variation  in path length through the particle bed.

A high speed grayscale camera (Point Grey BlackFly 12A2M) mounted directly beneath the output image port of the X-ray image intensifier (see schematic in Extended Data Fig.\,3) provides a subsurface bottom view of the tumbler. Images were acquired prior to each tumbling action: the first frame, when the free surface was horizontal and subsequent frames while the tumbler was slowly rotating to the angle of repose. The direction and magnitude of tracer particle displacement in these frames identified whether the particle was at the free surface or in the bulk near the tumbler wall. The image distortion due to image intensifier and camera optics was corrected using the \textsc{MATLAB} Image Processing Toolbox function \texttt{imwarp}. The geometric transformation required by the function \texttt{imwarp} was a 4th order polynomial model generated from applying the function \texttt{fitgeotrans} to a Cartesian hole pattern (diameter 3.5\,mm with 5.1\,mm spacing) image. The corrected hole pattern image was used to generate the matrix transformation from the hole pattern's pixel spacing to the physical grid dimensions. To track the tracer particle and the fiducial indicator, each image was divided by a background image generated from the average of all iterations from a particular run to reveal the two features of interest. The tracer particle and the fiducial indicator were distinguishable from each other by their eccentricity and were tracked automatically using 2D feature finding \textsc{MATLAB} algorithms developed by the Kilfoil group\footnote{The repository for the code is hosted at \url{http://people.umass.edu/kilfoil/tools.php}} using methods from Crocker et al.\cite{CrockerJColloid1996}

\subsection*{Flowing layer (FL) model}
The flowing layer (FL) model~\cite{MeierAdvPhys2007,ChristovSIAMJ2014} assumes that the flow is primarily two-dimensional in the streamwise direction and confined to a thin lenticular flowing layer with a constant depthwise shear rate $\dot{\gamma}$ for each tumbling action $\mathbf{U}$ and $\mathbf{W}$ (Extended Data Fig.\,2). Using a Cartesian coordinate system with its origin at the center of radius $R$ spherical tumbler, rotation is clockwise about the $z$-axis and $x$-axis at a rotation speed $\omega$ for an angular displacement $\theta_z$ and $\theta_x$ for $\mathbf{U}$ and $\mathbf{W}$ actions, respectively. For convenience, all variables are dimensionless---length scales ($x$, $y$, $z$, $r$, and flowing layer thickness $\delta$) are normalized by $R$ and rotation period $T$ is normalized by $1/\omega$. The interface between the flowing layer and the bulk is given by $\delta(x,z) = \epsilon\sqrt{1-x^2-z^2}$, where $\epsilon = \sqrt{\omega/\dot{\gamma}}$ is the maximal dimensionless flowing layer thickness (at the center $x=z=0$). The velocity $\mathbf{u} = (u,v,w)$ is piecewise defined for the flowing layer and the bulk. For the $\mathbf{U}$ action, the flowing layer $(y\geq-\delta)$ velocity is $\mathbf{u}_{\mathrm{fl}} = ((\delta+y)/\epsilon^2,xy/\delta,0)$ and the bulk $(y<-\delta)$ is in solid body rotation with velocity profile $\mathbf{u}_{\mathrm{b}} = (y,-x,0)$. Analogously, the velocity field for the $\mathbf{W}$ action is obtained by interchanging $x$- and $z$-components. Particle positions are generated by integrating the velocity field using the Runge-Kutta (RK4) method in the flowing layer and the semi-implicit Euler method in the bulk.\cite{MeierAdvPhys2007,ChristovSIAMJ2014} The dimensionless time step is $\Delta t/T = 5\cdot10^{-5}$ to ensure stability.  A flowing layer thickness at the midpoint of the flowing layer ($\delta(0,0)$) of $\epsilon = 0.15$ matches the experiments ($\omega$ = 2.6\,rpm) based on the tracer particle flowing layer passage time. To investigate the impact of the flowing layer thickness, conditions with $0 \le \epsilon \le 0.20$ were also simulated. For context, previous studies on quasi-2D flows reported flowing layer thickness of $\epsilon \approx 0.1$,\cite{FelixEPJE2007} depending on the particle to tumbler diameter ratio $d/D$.

Stroboscopic maps (also known as Poincar\'e sections or discrete time maps) of 500 iterations were used to investigate mixing and non-mixing behavior and to provide direct comparison to experiments. The stroboscopic maps utilized tracers seeded at the intersection of the flowing layer boundary and the $r=0.9$ hemispherical shell at zero iterations (blue points) and at the first half iteration (red points). Regions avoided by the tracer particles correspond to elliptic regions. The outer boundary orbits of these elliptic regions were obtained by finely seeding initial conditions near the boundary in successive stroboscopic maps until the boundary orbits were extracted. The periodicity of these elliptic regions was determined by tracking tracers seeded in these regions. Other larger elliptic orbits that wrap around the bulk and the flowing layer (e.g., grey points in Fig.~\ref{fig:protocol4545}(a-b) and in Extended Data Fig.~1(a)) were extracted by seeding a tracer in the neighbourhood of the orbit.

\subsection*{Piecewise Isometry (PWI) model.} The PWI model is applicable in the infinitely thin flowing layer (ITFL) limit ($\epsilon = 0$) of the FL model.\cite{JuarezEPL2010,JuarezChristovCES2012,ParkChaos2016} Like the continuum model, stroboscopic maps were used to investigate mixing and non-mixing regions. Because the flow dynamics are radially invariant with an ITFL, tracer positions are mapped by their angular displacements on the hemispherical shell. When tracers reach the free surface, they are instantaneously reflected across the ITFL in the streamwise direction. The initial positions of the passive tracers in the stroboscopic maps were selected to lie along the isometry partitions at zero iterations (blue points) and the first half iteration (red points) and mapped for 10,000 iterations to assure convergence.
}

\subsubsection*{SI Movies}


Video 1 $ \mid $ \textbf{Motion of mixing barriers under the ($45^{\circ}, 45^{\circ}$) protocol}. At finite flowing layer thickness ($\epsilon$ = 0.15), \textbf{a} bottom view of the bed, and \textbf{b} bottom view of the flowing layer. The initial condition in the video is generated by tracking the positions of particles, initially seeded on the edge of the flowing layer before the first and second rotations, for 500 iterations. Particles are advected for three full iterations in the video. The three blue ``fingers'' appear sequentially in the flowing layer at each full-iteration and return to their original locations after three iterations. Identical behavior is demonstrated by red ``fingers,'' but on half iterations.

Video 2 $ \mid $ \textbf{Motion of non-mixing regions under the ($57^{\circ}, 57^{\circ}$) protocol}. At finite flowing layer thickness ($\epsilon$ = 0.15), \textbf{a} bottom view of the bed, and \textbf{b} bottom view of the flowing layer. The initial condition in the video is generated by tracking the positions of particles, initially seeded on the edge of the flowing layer before the first and second rotations, for 500 iterations. Particles are advected for three full iterations in the video. During each rotation, a pair of blue and red open regions passes through the flowing layer and reappears in the bed on the other side after rotation completes. No open regions are present in the flowing layer when the rotation axis is switched.

\showmatmethods 

\acknow{Z.Z.~thanks D.~Diaz, M.~Krotter, and A.~Spitulnik for assistance with building the experimental apparatus. This research was funded by NSF Grant CMMI-1435065.}

\showacknow 

\pnasbreak

\bibliography{bib_BSTflow}

\end{document}